\documentclass[preprint,superscriptaddress]{revtex4}
\usepackage[dvips]{graphicx}
\usepackage{setspace}
\usepackage{natbib}
\usepackage{amssymb,amsfonts,amsmath}

\begin{document}
\begin{center}
{\large \bf
Intra- and intercellular fluctuations in Min protein dynamics decrease with cell length
\\}
\vspace{0.5cm}
Elisabeth Fischer-Friedrich$^{1}$, Giovanni Meacci$^{2}$, Joe Lutkenhaus$^{3}$, Hugues Chat\'e$^{4}$, and Karsten Kruse$^{5}$\\
\vspace{0.5cm}
\end{center}
$^1$\small{Max Planck Institute for the Physics of Complex Systems, N\"othnitzer Stra{\ss}e 38, 01187 Dresden, Germany.
Present address: Department of Physical Chemistry, Weizmann Institute of Science, Rehovot 76100, Israel}\\
$^2$\small{IBM T.~J. Watson Research Center, P.O. Box 218, Yorktown Heights, NY 10598.
Present address: Department of Biological Sciences, Columbia University, New York, NY 10027}\\
$^3$\small{Department of Microbiology, Molecular Genetics, and Immunology, University of Kansas Medical Center,
Kansas City, Kansas 66160}\\
$^4$\small{CEA-Saclay, Service de Physique de l'Etat Condens\'e, 91191 Gif-sur-Yvette, France}\\
$^5$\small{Theoretische Physik, Universit\"at des Saarlandes, Postfach 151150, 66041 Saarbr\"ucken, Germany}
\vspace{0.5cm}

\begin{center}
{\bf Abstract}
\end{center}
Self-organization of proteins in space and time is of crucial importance for the functioning
of cellular processes. Often, this organization takes place in the presence of strong random fluctuations due to the
small number of molecules involved. We report on stochastic switching of the Min-protein distributions between
the two cell-halves in short \textit{Escherichia coli} cells. A computational model provides strong evidence that
the macroscopic switching is rooted in microscopic noise on the molecular scale. In
longer bacteria, the switching turns into regular oscillations that are required for positioning of the division plane.
As the pattern becomes more regular, cell-to-cell variability also lessens, indicating cell length-dependent
regulation of Min-protein activity.\\
\\
{\bf keywords}: cell division, spatiotemporal patterns, fluctuations, protein self-organization\\
\newpage

\section{Introduction}
Subcellular structures are often formed by a small number of proteins. This is notably the case in prokaryotes,
as exemplified by the Soj proteins in \textit{Bacillus subtilis} that switch randomly between the two distal parts
of the nucleoid~\cite{quis99,mars99}, by the Min proteins in \textit{Escherichia coli}~\cite{lutk07} that switch
periodically between the two halves of the bacterium, as well as by the rings and helices formed by FtsZ and MreB
in many bacteria~\cite{pogl08}. The concentrations of these proteins are in the
micromolar range, hence, the above structures, that may extend over several micrometers,
 are formed by a few
hundred molecules. Such small numbers can imply large random fluctuations in space and
time~\cite{quis99,mars99}.

However, there is currently a lack of quantitative experimental studies of
random fluctuations in spatially extended protein
structures. In particular it remains largely unexplored how fluctuations on cellular scales originate from the interplay of molecular
noise, i.e., the stochastic nature of the protein kinetics, and external noise, i.e., variations in the environment or in
the number of
proteins present in the cell~\cite{howa03,doub05,kerr06,tost06,fang06}. In the context of gene expression, random
fluctuations contribute essentially to the dynamics~\cite{raj08}. In particular, they directly lead to differences
in genetically identical cells~\cite{elow02}, i.e., intercellular fluctuations, and can play an important role in cell
fate decisions~\cite{elow02, losi08}. In this context, cell size can act as a control parameter to regulate either
the amplitude of fluctuations~\cite{suel07} or the distribution of phenotypes they generate~\cite{suel07, stpi08}.

In this work, we report on intra- and intercellular fluctuations in the spatiotemporal patterns formed by the Min
proteins in \textit{E.~coli}. These proteins select the cell center as the site of division~\cite{lutk07}. Their distribution
has been found to change periodically with time, such that most of the proteins reside for about 40s in one
cell half and subsequently for the same time in the opposite cell half~\cite{rask99}. As one of the Min proteins, MinC,
inhibits formation of the Z-ring, division is in this way suppressed at the cell poles. Computational analysis of
the Min-protein dynamics indicate that the observed pattern results from self-organization of the ATPase MinD
and MinE~\cite{krus07}. This idea is
supported by experiments {\it in vitro} in which planar and spiral waves of MinD and MinE emerged
spontaneously in the presence of ATP on a supported lipid bilayer~\cite{loos08}.

We found, that in cells below
a critical length of 2.7$\mu$m the Min distribution switches stochastically between the two cell halves. In this
phase intercellular fluctuations are pronounced. In cells longer than the critical length, we observed regular
oscillations and intercellular fluctuations were significantly reduced. Computational analysis shows that
stochastic switching can result from molecular noise. Furthermore, it indicates that, in the stochastic state, the
rate of ATP consumption per MinD molecule is only half of that in the oscillatory state.

\section{Results}
\begin{figure}
   \begin{center}
      \includegraphics*[angle=0,width=5.25in]{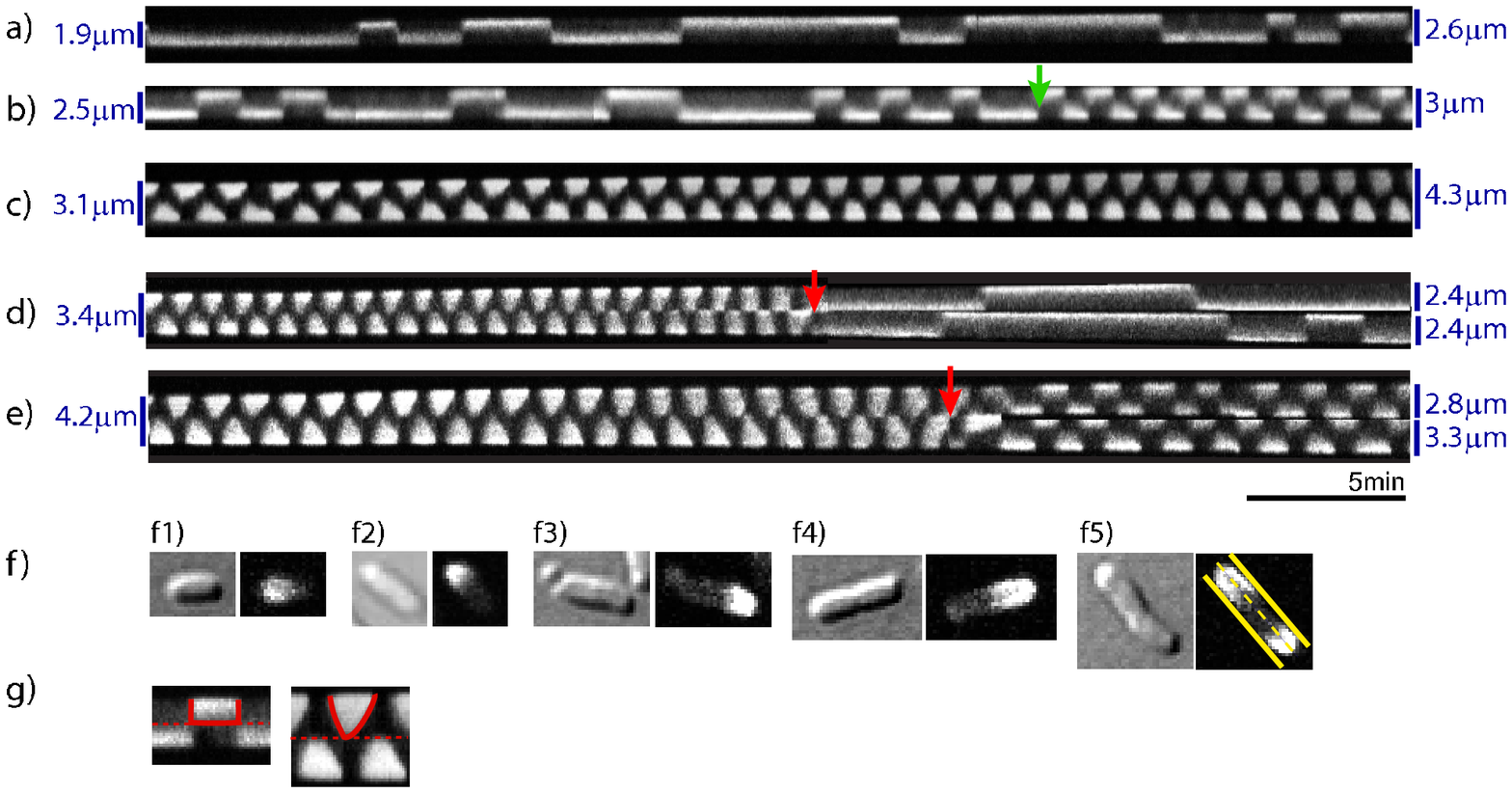}
\caption{Distribution of MinD-GFP along the bacterial long axis as a function of time.
Time increases to the right, the long axes are oriented vertically.
Initial cell lengths are given on the left, final cell lengths on the right.
a) Stochastic exchange of MinD between the two cell halves.
b) Stochastic exchange turned into regular oscillations at a length of $2.8\mu$m
(green arrow).
c) Regular Min oscillations with a period of $73{\rm s}$.
d, e) Min dynamics in dividing cells. Daughter cells show either stochastic switching (d) or regular oscillations
(e). Divisions occurred, respectively, at lengths of 4$\mu$m and 5$\mu$m of the mother cells (red arrows).
f) Nomarsky and fluorescence images of the cells used in (a)-(e) at the beginning of the measurements.
In (f5), lines indicate the bacterial long axis and width.
g) Extracts of the kymographs shown in a) and c), respectively.}
      \label{fig1}
   \end{center}
\end{figure}

We observed the distribution of MinD for 40 minutes in each of 209 cells, see Materials and Methods, during
which individual bacteria grew roughly 0.5$\mu$m. In cells shorter than 2.5$\mu$m, instead of oscillating regularly,
MinD typically shifted stochastically from one cell half to the other, see Fig.~\ref{fig1}a. The residence times
of MinD in one cell half varied widely in these cells, whereas
complete switching from one cell half to the other occurred
in an interval of less than 15s. The extent of the region covered by MinD in one cell half did not change
notably between two exchange events, see Fig.~\ref{fig1}g.

For cell lengths between $2.5\mu$m and 3$\mu{\rm m}$, the Min pattern typically changed from stochastic
switching to regular oscillations with a period of about 80s, see Fig.~\ref{fig1}b and Fig.~S1 in the Supp.~Mat. The precise lengths at which
this transition occurred differed between cells. In the oscillatory regime, between two switching events, the
region covered by MinD first grew monotonically from the cell end and then shrank monotonically towards
the same end, see Fig.~\ref{fig1}g. Analogously to the stochastic exchange regime,
the transition of MinD from one half to the other was fast compared to the oscillation period. In a small number
of cells (N=5), we observed a transition back from regular oscillations to stochastic switching.

Cells longer than $3.5\mu{\rm m}$ invariably displayed regular oscillations, see Fig.~\ref{fig1}c and Fig.~S1 in the
Supp.~Mat. The oscillation
period typically decreased slightly with increasing cell length. For the cell shown in Fig.~\ref{fig1}c, the initial
period is approximately $87$s, while it is approximately $70$s at the end. Similar behavior can be observed
for MinE, see Supplementary Material.

About 5\% of the cells divided during the observation time, see Fig.~\ref{fig1}d, e.
In all these cases, MinD oscillated regularly prior to division. Consistent with our findings in non-dividing cells,
the pattern displayed by the Min proteins in the daughter cells immediately after division correlated with their
length: daughter cells shorter than 2.5$\mu$m typically displayed stochastic switching, while the Min proteins
mostly oscillated in daughter cells longer than 3$\mu$m. In some cases, however, the two daughter cells
showed different MinD patterns, in spite of having equal lengths: while in one daughter we observed oscillations,
the other displayed stochastic shifts. In these cases, judged from the fluorescence intensities,
the distribution of MinD between the two daughter cells was significantly uneven \footnote{In some cases,
the oscillatory cell was brighter, in others, the stochastically switching cell. This suggests that
the protein number is not the essential difference between oscillating and stochastically switching cells.}.

\subsection{Distribution of residence times} The transition from stochastic switching to regular oscillations
is accompanied by a qualitative change in the
distribution of residence times $\tau$ cumulated over cells of a given length,
see Fig.~\ref{fig2}. For cells longer than 3$\mu$m,
this distribution shows a pronounced peak at $\approx 35s$, which essentially
coincides with the mean value, and a small
tail including rare events of residence times of up to $200s$.
These rare events correspond to instances when one
or more oscillations
were missed and thus the residence time extends to values significantly
larger than the mean, see Fig.~\ref{supp3} in the Supplementary Material.
The left, main part of the distribution ($\tau<60\rm s$)
can be fitted  by a log-normal distribution with a geometric mean of  $37.9{\rm s}$
and a geometric standard deviation of $1.2{\rm s}$.
The tail ($\tau>60\rm s$) can be fitted
to an
algebraic decay with  decay exponent $\alpha=3.6\pm 0.6$.

For cells smaller than 2.5$\mu$m, the distribution of
residence times is qualitatively different: it is essentially algebraic,
$\propto\tau^{-\alpha}$, with a decay exponent $\alpha=2.1\pm 0.2$.
Consequently, the mean-value is just at the
edge of being well-defined, while the variance diverges,
implying enormous variations. This is in
contrast to a usual random telegraph process in which a system
switches stochastically between two states at given constant rates.
In this case, the distribution of residence times in one state decays
exponentially. We, therefore, also fitted an exponential decay to the
tail of the distribution of residence times. However, the logarithm of the likelihood ratio of an
algebraic to an exponential decay was 35 and thus strongly indicative
of an algebraic decay~\cite{clau07}.

In  Fig.~\ref{fig3}a, we show the functional dependence on the cell length of the mean residence time
$\langle\tau\rangle$ cumulated over all events and all cells in a given length interval. Cell lengths
were binned into intervals of $0.2\mu {\rm m}$ length and residence times were assigned to the length
of the cell at the beginning of the respective residence time interval. Note, that since within 500s,
which is exceptionally large for a residence time, the bacteria grew at most 0.2$\mu$m in length, our results do not
depend significantly on the exact assignment rule. Strikingly, the variation of  $\langle\tau\rangle$ is
well-described by two exponential decreases: For cell lengths smaller than 2.7$\mu$m that are typically
in the stochastic regime,  $\langle\tau\rangle \propto\exp(-L/\lambda)$ with $\lambda=0.56\mu$m,
while for longer, typically oscillating cells $\lambda=2\mu$m.

To characterize fluctuations around $\langle\tau\rangle$, we computed the effective standard deviation
$\sigma$ corresponding to our finite number of samples \footnote{Even though the standard deviation
would be infinite if the functional form of the distribution extended over all residence times, our finite
samples which have roughly the same size for each length interval, allow for a meaningful estimate of
$\sigma$.}. The effective
standard deviation $\sigma$, too, decreases exponentially with increasing cell length. Note that, for
cells smaller than approximately 2.7$\mu$m it is larger than $\langle\tau\rangle$, while for longer cells
it is smaller. For cells larger than 3.5$\mu$m, the standard deviation
is below 10s indicating that the oscillation period varies remarkably
little within a single cell and within an ensemble of cells.

\subsection{Cell-cell variability} To assess the contribution of cell-to-cell variability to the fluctuations, we compared the mean residence
time $\bar\tau_{\rm ic}$ of individual cells with that of other cells sharing the same cell length.
In Fig.~\ref{fig3}b, we represent the normalized standard deviation of the distribution of mean residence
times,  $\sigma_{\rm ic}=\sqrt{\langle\bar\tau_{\rm ic}^2\rangle-\langle\bar\tau_{\rm ic}\rangle^2}/
\langle\bar\tau_{\rm ic}\rangle$. It decreases with increasing cell length, first mildly then sharply
for cell lengths larger than 3$\mu$m, indicating a clear reduction of cell-to-cell variability in longer cells.
This suggests a control mechanism that adjusts Min parameters towards optimal reference values when
the cells grow.

\subsection{Mathematical modeling}
Several attempts at mathematically modeling of the Min system have been successful in reproducing essential aspects of the  Min
dynamics~\cite{krus07}. Based on different underlying microscopic pictures, these models uniformly explain the Min oscillations as
an emergent property
of interacting MinD and MinE in presence of a membrane and ATP. In particular, self-enhanced binding of MinD and/or MinE to the membrane, or aggregation of membrane-bound proteins can trigger an instability towards oscillations. Biochemically, MinD and MinE are not yet characterized well enough to exclude one possibility or the other. The validity of a model can alternatively be tested by verifying
predictions of the macroscopic behavior. For example, the aggregation models in Refs.~\cite{krus02,meac05} predict the existence
of stationary heterogeneous protein distributions.

We tested several stochastic versions of models presented in the past~\cite{huan03,meac05,loos08} and checked if they showed stochastic switching  of Min proteins in certain parameter regimes, see Supp. Mat.
Out of these, only  the model suggested in \cite{meac05} produced stochastic switching similar to what is observed in the
cell \footnote{Also in the deterministic model of Ref.~\cite{mein01} bistable stationary states
can be observed (H. Meinhardt, personal communication).
However, the model relies essentially on protein synthesis and degradation, which
have been shown to be irrelevant for the Min oscillations.}. Since stochastic simulations in three dimensions yield essentially
the same results as in one dimension, cf.~Refs.~\cite{kerr06} and \cite{fang06}, we restricted our attention to simulations in one
dimension.

What is at the origin of the transition between stochastic switching and oscillations of MinD and the
corresponding change in the residence-time distribution?
The experimental data presented above show that the spatiotemporal
structure of the MinD pattern correlates with cell length. As a naive guess
one might think that the cell length directly controls the pattern. However, in our experiments, we occasionally
observed that two daughter cells of about the same length showed
different MinD patterns right after division of the mother cell. In addition,
in some instances, cells shorter than 2.5$\mu$m also showed oscillations,
while in some cells longer than 3$\mu$m MinD switched stochastically
between the two cell halves. This suggests that, in addition to the cell length, other factors
 control  the transition from stochastic switching to regular oscillations.

In order to better understand which parameters of the system are able to trigger stochasticity in Min switching, we used a one-dimensional particle-based stochastic version of the one model which we identified as being able to produce stochastic switching, see Fig.~\ref{fig4} and Material and Methods.
We find that, increasing the cell length
is not sufficient to generate a transition from stochastic switching to oscillatory Min dynamics if, simultaneously, protein numbers increase
proportionally to the cell length.
 Changing other parameters of the Min dynamics, however, can trigger such a transition. As an example we now discuss
alterations in the rate $\omega_E$ of MinE binding to the membrane,
see  Fig.~\ref{fig5}.  Let us stress that in the following we keep the total protein concentrations constant, that is, protein
numbers vary proportionally to the cell length.
For $\omega_E=0.1s^{-1}$ and a cell length of $2\mu$m, our simulations show stochastic switching, while
for $\omega_E=0.35s^{-1}$ and a cell length of $3\mu$m, we find oscillations \footnote{A similar transition is found
for a fixed system length of $2\mu$m. Furthermore, the transition can be induced also by varying
the particle numbers or other rates appropriately (see also Supp. Material).},
see Fig.~\ref{fig5}a.
The model is thus capable of reproducing the observed transition by changing
MinE activity. It is important to note, that this transition is not
a consequence of approaching the deterministic limit, but is
inherent to the Min-protein dynamics. Indeed, for the respective binding
rates, the deterministic model is either bistable and settling into one of
two mirror-symmetric stationary states
or it generates oscillations~\cite{krus02,meac05}.  We conclude that in short cells
the macroscopic switching behavior
is rooted in intrinsic fluctuations of the molecular processes
due to the relatively small number of proteins involved.
The transition from stochastic switching to oscillations, however, is not due to the discreteness
of molecule numbers, which in general is possible~\cite{toga01}.

\begin{figure}
   \begin{center}
      \includegraphics*[angle=0,width=5.25in]{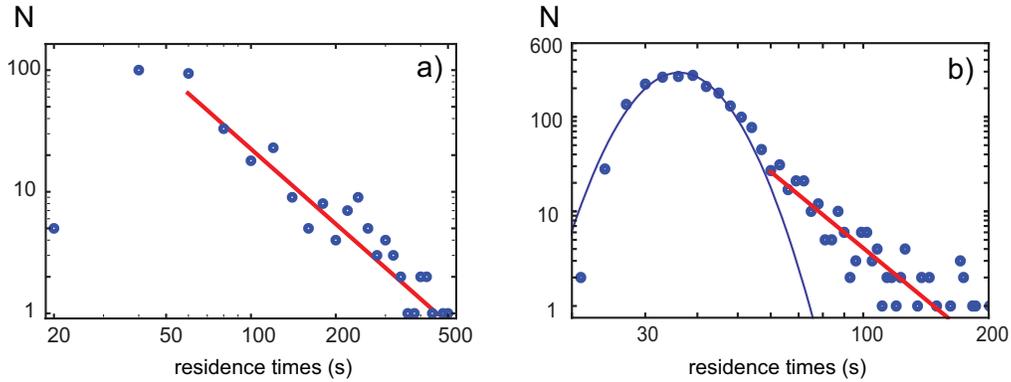}
\caption{Ensemble distributions of residence times.  a) Residence times for initial cell lengths between 2 and
2.2$\mu$m (total: 358) and b) for initial lengths greater than 3$\mu$m (total: 2157). Times are respectively binned into $20{\rm s}$  (a)
and into $3{\rm s}$ intervals (b). Solid lines result from fits of the distribution tail ($\tau<60s$). (a) Fitting of an
algebraic tail yields an exponent $-2.1\pm 0.2$ (red line).
(b) the main part of the distribution is well fitted to a log-normal
distribution with geometric mean $37.5{\rm s}$ and geometric
standard deviation $1.2{\rm s}$ (blue line); the tail of rare events of
duration longer than 60s is reasonably well fitted to an algebraic law with
exponent $-3.6\pm 0.6$ (red line).}
      \label{fig2}
   \end{center}
\end{figure}

\begin{figure}
   \begin{center}
      \includegraphics*[angle=0,width=5.25in]{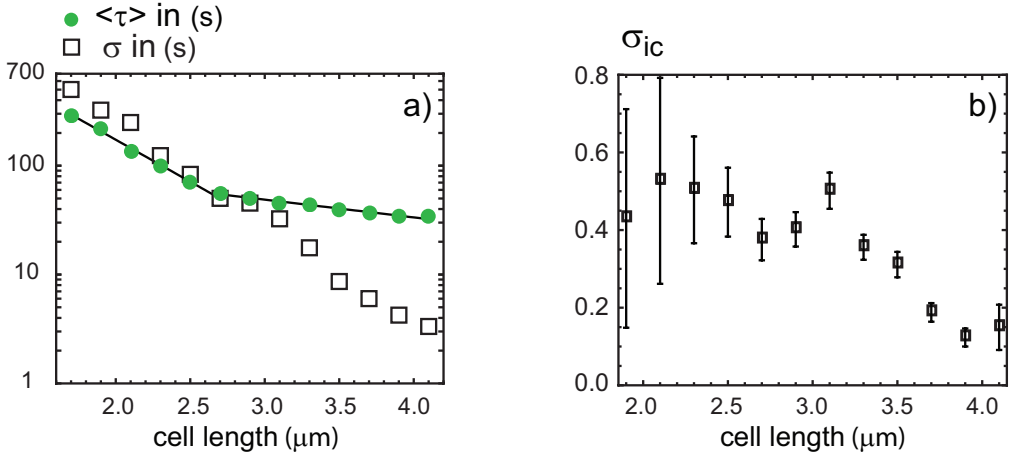}
\caption{a) Mean  $\langle\tau\rangle$ and effective standard deviation
$\sigma$ of ensemble distribution of residence times
as a function of cell length $L$.
Cell lengths have been binned into $0.2\mu {\rm m}$
intervals. Straight lines are exponential fits to the
$\langle\tau\rangle$ data, $\tau = \tau_0\exp(-L/\lambda)$.
For cell lengths below $2.7\mu$m we find $\lambda=0.5\mu$m, for larger
lengths $\lambda=2\mu$m.
b) Relative standard deviation $\sigma_{\rm ic}=
\sqrt{\langle\bar\tau_{\rm ic}^2\rangle-\langle\bar\tau_{\rm ic}\rangle^2}/
\langle\bar\tau_{\rm ic}\rangle$ of the distribution of individual cell
mean residence times $\tau_{\rm ic}$ as a function of cell length.}
      \label{fig3}
   \end{center}
\end{figure}

\begin{figure}
   \begin{center}
      \includegraphics*[angle=0,width=5.25in]{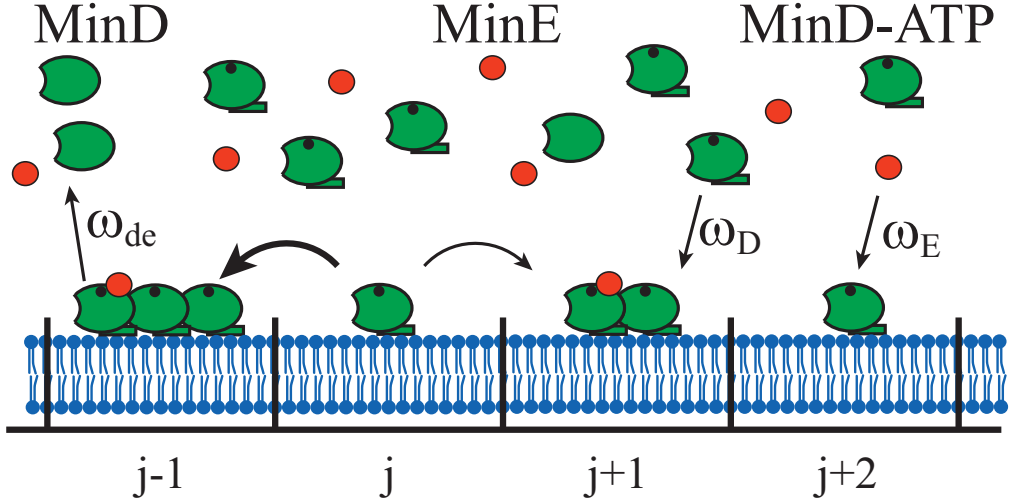}
\caption{Stochastic model of Min-protein dynamics. The cell membrane is represented by a one-dimensional array
of bins that can take up MinD-ATP (green) and MinE (red). Cytosolic MinD-ATP binds at rate $\omega_D$, cytosolic
MinE binds to membrane-bound MinD-ATP at rate $\omega_E$, MinD-ATP-MinE complexes detach at rate $\omega_{de}$.
The distribution of proteins in the cytosol is assumed to be homogenous \cite{meac06} and the exchange of MinD-ADP to MinD-ATP
infinitely fast. On the membrane MinD-ATP diffuses and is attracted towards other MinD-ATP. For details see Materials
and Methods.}
      \label{fig4}
   \end{center}
\end{figure}

Quantitative analysis of simulations corresponding to a single cell in
the stochastic switching state, presented in Fig~\ref{fig5}a, yield an
exponential decay of the residence times as expected for a random telegraph process.
This is however {\it not} in contradiction
with the experimental data, for which {\it ensemble} averaging over different cells yields
algebraic decay, Fig.~\ref{fig2}a. Indeed, experiments revealed
cell-to-cell variability, particularly strong amid short cells exhibiting stochastic switching,
Fig.~\ref{fig3}b. As we show below, introducing such variability in the model resolves
the apparent contradiction.
It is conceivable that in the experiments the cell-to-cell variations in the Min dynamics
for a given cell length result mainly
from differences in the numbers of MinD and MinE.
Indeed,  we found in $\approx 65\%$ of the observed divisions that the
numbers in the daughter cells, as judged from
fluorescence intensities, differed by more than 10\%,
see Supplementary Material.

In order to test the effects of varying protein numbers in the model,
we performed several simulation runs with
MinD protein numbers drawn at random from a Gaussian distribution with a standard deviation of
10\% of the mean~\cite{elow02}. The ratio of MinD to MinE was fixed
to 8/3~\cite{shih02} and all other parameters kept constant.
Lumping together the observed residence times
of 70 runs with different protein numbers, we now observe an
algebraic decay with exponent $2.03\pm 0.2$, see Fig.~\ref{fig5}b,
which is very close to the experimental value of 2.1, see Fig.~\ref{fig2}a.
As in the experiments, the mean is barely defined and the standard deviation
diverges. This result constitutes a second instance of the general finding
that power law distributions can result from Gaussian variations of
system parameters as discussed by Tu and Grinstein in the context
of the bacterial flagellar motor~\cite{tu05}.

Analogously, we calculated the distribution of residence times in the oscillatory
regime, see Fig.~\ref{fig5}c. It is
remarkably similar to that found experimentally, see Fig.~\ref{fig2}b:
a large hump, well fitted to a log normal distribution with geometric mean 31s and
geometric standard deviation 1.2s, plus a small tail at large residence times which
can be fitted by an algebraic decay with exponent $\alpha=4.6\pm0.4$, similar to
the experimental value $\alpha=3.6$. Moreover, an inspection of the rare
events at the origin of this algebraic tail showed that they are instances of
``missed'' oscillations, just like in the experimental kymographs, see Fig.~\ref{supp3}
in the Supplementary Material. We conclude that the oscillation period is
a robust property of the Min System with respect to fluctuations in protein numbers.

We then investigated the transition from the switching to the oscillatory state
in more detail. Fig.~\ref{fig5}d presents the mean value and the standard
deviation as a function of system length. The rate $\omega_E$ was varied
with the system length according to a sigmoidal dependence,
see Fig.~\ref{fig5}e inset and Supplementary Material. Similar to the experimental results, we
find a decreasing mean and standard deviation with increasing length and
the transition from stochastic switching to oscillations is accompanied by
a drop of the standard deviation below the mean value.
The simple particle
model of the Min dynamics thus semi-quantitatively captures the effects of
length changes and of fluctuations in the system.

Finally, we used the computational model to infer the ATP consumption rate as a function
of cell length. Interestingly,
the average ATP consumption rate per unit length, which is proportional to the ATP consumption rate
per MinD, increases with cell length until the transition
from stochastic switching to regular oscillations and then remains roughly constant, see Fig.~\ref{fig5}e.
Our calculations thus let us hypothesize that is energetically advantageous for a short
cell to keep the Min system in the
stochastic switching regime.

\section{Discussion}
In summary, we have shown that in short \textit{E. coli} cells the distribution of MinD proteins stochastically
switches between two mirror-symmetric states, while it oscillates regularly in longer cells. The Min system provides
thus an intriguing example of a spatiotemporal pattern under physiological conditions that combines regularity with
stochastic elements.

Stochastic switching between two states is also known for the Soj/Spo0J proteins which in \textit{B.~subtilis}
relocates irregularly between the two sides of the bacterial nucleoid~\cite{quis99,mars99}. A computational
model suggests that this phenomenon, too, is rooted in molecular fluctuations~\cite{doub05}.
Similarities between MinD and Soj had already been pointed out due to
resemblance in structure, polymerization and ATPase
activity~\cite{loew09}. However, notable differences exist as, for example, Spo0J seems to be
always bound to the nucleoid unlike MinE that detaches from the membrane. Furthermore, no
oscillations have been found for Soj/Spo0J.

Based on our theoretical analysis, we can speculate about a possible regulatory mechanism controlling
the dynamics of the Min proteins in \textit{E.~coli}.
We interpret the findings of our computations as indications
of changes in the activity of the
Min proteins as the cells grow. Specifically, the ability of MinE to bind to membrane bound MinD
might be reduced in early phases of the cell cycle.  According to
this view, the Min system is kept in a stand-by mode with reduced ATP consumption  in short cells and gets fully activated
only as a cell approaches division, when the Min system needs to be functional. Why does stochastic switching of the Min
proteins not lead to minicelling? The central Z-ring needs about 5 to 10 minutes to mature and
polar Z-rings are less stable than the central one. Consequently, polar rings formed in an early
phase of the cell cycle might be disassembled by MinC after oscillations have set in, which occurs
significantly before septation starts.

Obviously, the transition might be also due
to other evolutionary constraints or be just a side-effect of the Min-protein dynamics. Independently of the
transition's physiological reason, we propose
 that changing system parameters such that the dynamic behavior is qualitatively modified
through a bifurcation - in the present case from bistable to oscillatory - presents an interesting generic
mechanism to tame detrimental fluctuations
as subcellular processes become vital.

\section*{Materials}
\subsection{Data acquisition} We used cells of the \textit{E.~coli} strain JS964 containing the plasmid
pAM238 encoding for MinE and GFP-MinD under the control of the lac-Promoter~\cite{hu99}.
Bacteria were grown overnight in a $3{\rm ml}$ LB medium at $37^\circ$C. 
Cells were induced with Isopropyl-$\beta$-D-thiogalactopyranosid (IPTG) at a concentration of 200$\mu$M and incubated for 4 hours.
During measurements, cells were in the exponential growth phase. The samples  were kept at a temperature
of $30^\circ$C using a Bachhoffer chamber. To keep bacteria from moving under the cover slip, we put them on an agar pad (1\% agar solution in LB medium with a reduced  yeast extract fraction, 10\%{},  in order to lower background fluorescence). The fluorescence recordings were taken with an Olympus FV
1000 confocal microscope, at an excitation wavelength of $488{\rm nm}$ from a helium laser at low power.
We used an Olympus UPLSAPO 60x, NA 1.35 oil immersion objective and recorded a frame every $3$s.
A measurement lasted 40min. During this period, the focus was manually readjusted at irregular intervals.
We were not able to determine the protein numbers in individual cells.

In total, we extracted data from 209 MinD-fluorescent cells obtained from 5 different measurements. We extracted data only
for cells which, at the beginning of the measurement, were smaller than
$3\mu {\rm m}$. Cell lengths were determined from differential interference contrast (DIC) images with an
accuracy of $\pm150{\rm nm}$ at the beginning and the end of a measurement. The cell length in-between
was determined by linear interpolation. Some of the cells in the field of view divided during the measurement
time. If division occurred after more than 20min of measurement, they were included in the data analysis. In these cases,
fluorescence recordings were used until cell constriction terminated.

To observe MinE distribution in cells, we used the strain WM1079 expressing MinD and MinE-GFP on a plasmid under the control of the pBAD-promoter \cite{corb02}. Bacteria were induced with Arabinose but otherwise grown in the same way as described above.

\subsection{Data analysis}
For a quantitative analysis, we mapped the time-lapse data of MinD-fluorescence from {\it E.~coli} cells onto the states of a two-state process. We transformed the fluorescence data  into a real-valued
time-dependent function by subtracting the integrated fluorescence intensities from the left and the right cell half, see
 Fig.~\ref{fig6}. Then, the moving
average over four time points was taken in order to reduce noise. The resultant function $f$ is positive when
the fluorescence maximum is in one cell half and negative in the opposite case. The residence time $\tau$ of MinD in one cell
half is defined as the interval between two consecutive zeros of $f$. To each residence time, we assigned
the cell length at the start of the respective residence period.

\subsection{Computational model of the Min dynamics} For the computational analysis of the Min-protein dynamics,
we use a particle-based version of the model suggested in~\cite{meac05}. It describes the formation of Min
oscillations on the basis of
an aggregation current of bound MinD that is generated by mutual attraction of the proteins. Furthermore,
it accounts for the exchange of MinD and MinE between the membrane and the cytoplasm, where MinE
only binds to membrane-bound MinD and where MinD detaches from the membrane only in the presence
of MinE. Measurements of the cytoplasmic diffusion constants of MinD and MinE
yielded values larger than 10$\mu$m$^2$/s~\cite{meac06}. We thus consider the limit of large cytoplasmic
diffusion which effectively leads to homogeneous cytoplasmic concentrations. The diffusion constant of
membrane-bound proteins is about two orders of magnitude smaller than for cytosolic diffusion~\cite{meac06}.
This mobility suffices to generate a sufficiently strong aggregation current generating an instability of the
homogenous protein distribution~\cite{meac05}.

A corresponding stochastic particle-based description is defined on a one-dimensional lattice with $N$ sites
representing the long axis of the cell.
The lattice spacing $l_b$ is chosen such that it is much larger than the protein size and much smaller than the
characteristic length of the Min pattern. Each site can contain at most $n_{\rm max}$ proteins, a number which can be understood as the circumference of the cell. We
assume diffusional mixing such that proteins are indistinguishable on a site. For site $j$,
the probability of attachment of cytoplasmic MinD and MinE during a sufficiently small time step $\Delta t$
is given by
\begin{eqnarray}
P_{D\rightarrow d}&=&\Delta t~ \omega_D \left(\frac{N_D}{N}\right)(1-\frac{n_{d,j}+n_{de,j}}{n_{\rm max}})\\
P_{E\rightarrow de}&=&\Delta t~ \omega_E\left(\frac{N_E}{N}\right)\frac{n_{d,j}}{n_{\rm max}}\quad,
\end{eqnarray}
respectively. Here, $\omega_D, \omega_E$ are the corresponding attachment rates and
$N_D$ and $N_E$ are, respectively, the numbers of cytoplasmic MinD and MinE. The numbers of membrane-bound
MinD and MinDE complexes on site $j$ are $n_{d,j}$ and $n_{de,j}$. The detachment probability is
\begin{equation}
P_{de \rightarrow E+D}=\Delta t~ \omega_{de} n_{de,j}\quad,
\end{equation}
where $\omega_{de}$ is the detachment rate. The exchange of particles between sites is governed by
\begin{equation}
P_{j\rightarrow j\pm 1}=\frac{D_d\Delta t}{l_b^2}n_{d,j}(1-\frac{n_{d,j\pm 1}+ n_{de,j\pm 1}}{n_{\rm max}})
I_{j\rightarrow j\pm 1}\quad,
\end{equation}
where
\begin{equation}
I_{j\rightarrow j\pm 1}=\left\{ \begin{array}{c@{\quad {\rm if} \quad}l}
1 & \Delta E_j<0 \\ \exp(-\frac{\Delta E_j}{k_B T}) & \Delta E_j>0
\end{array}\right.
\end{equation}
with $\Delta E_j=V_{j\pm 1}-V_j$. The potential $V$ describes the interaction strength between Min-proteins
on the membrane. We assume a square hole potential
\begin{equation}
V_j=-\frac{1}{n_{\rm max}}\left[ \frac{g_d}{(2R_d+1)}\sum_{j=-R_d}^{R_d}n_{d,j}+\frac{g_{de}}{(2R_{de}+1)}\sum_{j=-R_{de}}^{R_{de}} n_{de,j}\right].
\end{equation}
Here, the integers $R_d$ and $R_{de}$ relate to the ranges $r_d$ of the MinD-MinD interaction
and $r_{de}$ of the MinD-MinDE interaction through $R_d\simeq r_d/l_b$ and $R_{de}\simeq r_{de}/l_b$.
The parameters $g_d$ and $g_{de}$ tune the interaction strength. The diffusion constant of membrane-bound MinD
is $D_d$.

\begin{figure}
   \begin{center}
      \includegraphics*[angle=0,width=5.25in]{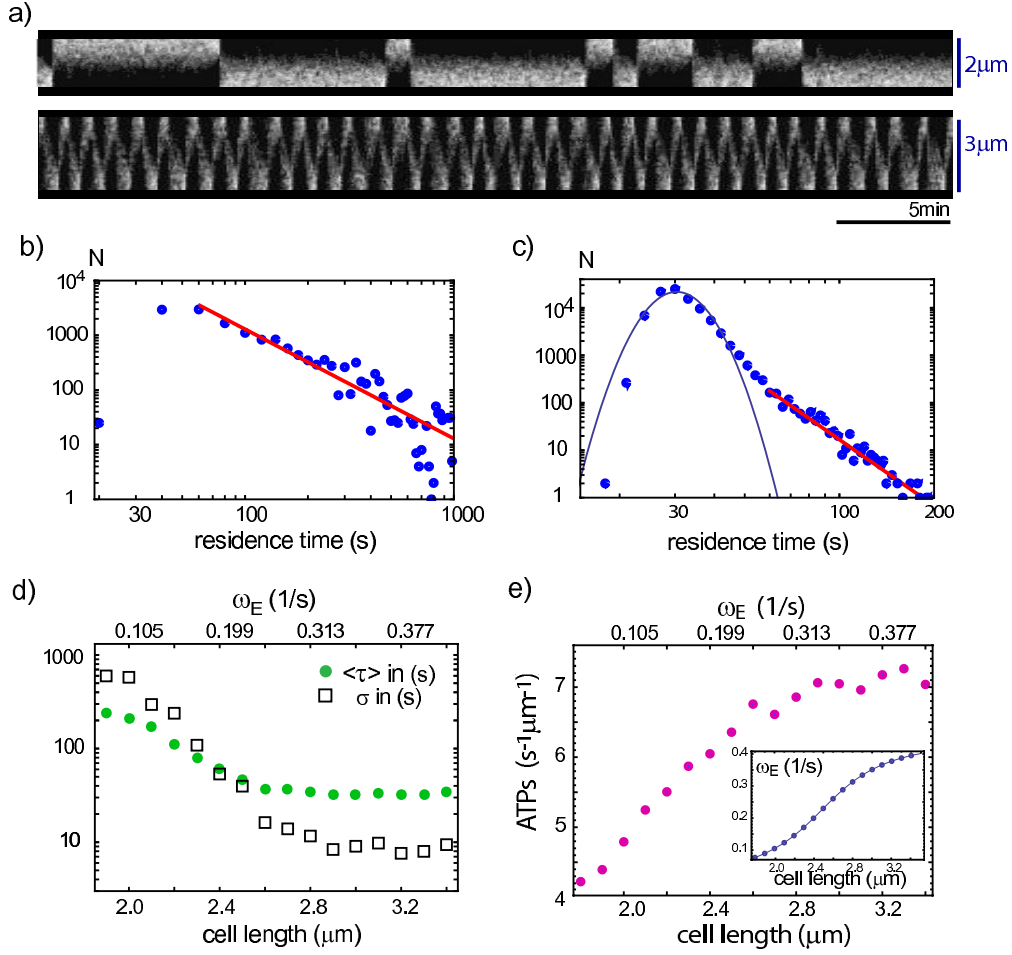}
\caption{Theoretical analysis of the Min-protein dynamics.
a) Kymographs obtained from simulations with $\omega_E=0.1{\rm s}^{-1}$ and length
$2\mu$m (top) and $\omega_E=0.35{\rm s}^{-1}$ and length $3\mu$m
(bottom), other parameters as below. Scale bar indicates 5min.
b,c) Ensemble distribution of residence times for the case of stochastic switching ($\omega_E=0.1{\rm s}^{-1}$, cell length: $2\mu$m) and oscillatory dynamics ($\omega_E=0.35{\rm s}^{-1}$,  cell length:$3\mu$m). Each histogram was obtained from 70 runs where the number
of MinD was drawn from Gaussian distributions with
mean 1440 (b) and 2160 (c), respectively and 10\% standard deviation.
The ratio of MinD to MinE proteins was fixed at 8/3.
The red lines represent algebraic fits $\tau^{-\alpha}$ for residence
times $\tau>60s$. For (b) $\alpha=2.03\pm 0.2$, for (c) $\alpha=4.6\pm0.4$.
In (c), the blue line indicates a log-normal distribution with geometric
mean $31{\rm s}$
and geometric standard deviation $1.2{\rm s}$.
d) Dependence of the mean residence time $\langle\tau\rangle$ (green circles)
and effective standard deviation $\sigma$ (black
squares) on the cell length obtained from ensemble simulations
as in (b) and (c).
e) The rate of ATP hydrolysis per unit length as a function of cell length. Inset:
dependence of the MinE attachment rate $\omega_E$ on cell length.
Other parameters are $\omega_D=0.04s^{-1}$,
$\omega_{de}=0.04s^{-1}$, $D_d=0.06\mu {\rm m}^2/s$, $r_d=1.2\mu{\rm
m}$, $r_{de}=0.1\mu{\rm m}$,
$g_d=35k_B T$, $g_{de}=-20k_B T$, $n_{\rm max}=43$ and bin length $l_b=33$nm.}
      \label{fig5}
   \end{center}
\end{figure}

\begin{figure}
   \begin{center}
      \includegraphics*[angle=0,width=5.25in]{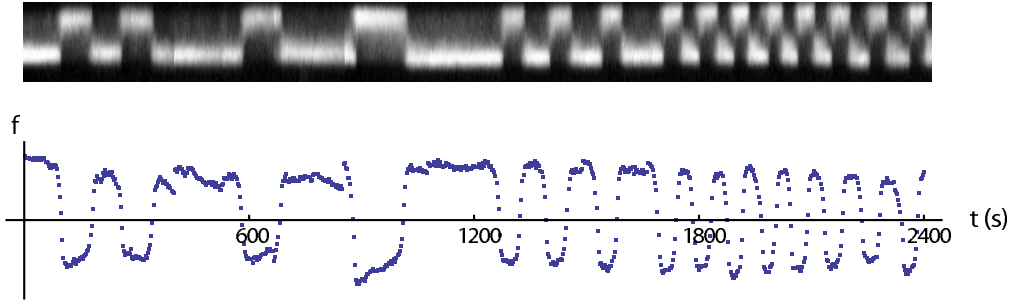}
\caption{Kymograph of GFP-MinD fluorescence and corresponding intensity curve $f$
obtained by subtracting the fluorescence intensity of one cell half from the intensity of the other cell half.}
      \label{fig6}
   \end{center}
\end{figure}

In Fig.~\ref{fig5}d, the attachment rate $\omega_E$ is increased jointly with the cell length in the simulations. We chose a sigmoidal increase of $\omega_E$ in dependence of the cell length $\ell_c$ according to the Goldbeter-Koshland function, which gives the
mole fraction of modified proteins that are under control of a modifying enzyme~\cite{gold81}
\begin{equation}
G(\ell_c)=\frac{\omega_E^{\rm sat} \times 2 \ell_c K}{vJ+\ell_c(K-1)+\sqrt{(vJ+\ell_c(K-1))^2-4(v-\ell_c)\ell_c K}}.
\end{equation}
We chose parameters $v=2.47, J=1.116$ and $K=0.099$. The factor $\omega_E^{\rm sat}$ determines the saturation value and was chosen as $0.4 {\rm s}^{-1}$. The precise functional form of $\omega_E(\ell_c)$, however, is not of importance as long as it is sigmoidal.


\begin{acknowledgments}
We thank M. B\"ar for useful discussions on the stochastic model.
\end{acknowledgments}



\clearpage

\newpage
\clearpage

\section{Supplementary Material}
\subsection{Stochasticity of Min Switching Decreases in Individual Cells During Cell Elongation/Aging}
\begin{figure}
   \begin{center}
      \includegraphics*[angle=0,width=4in]{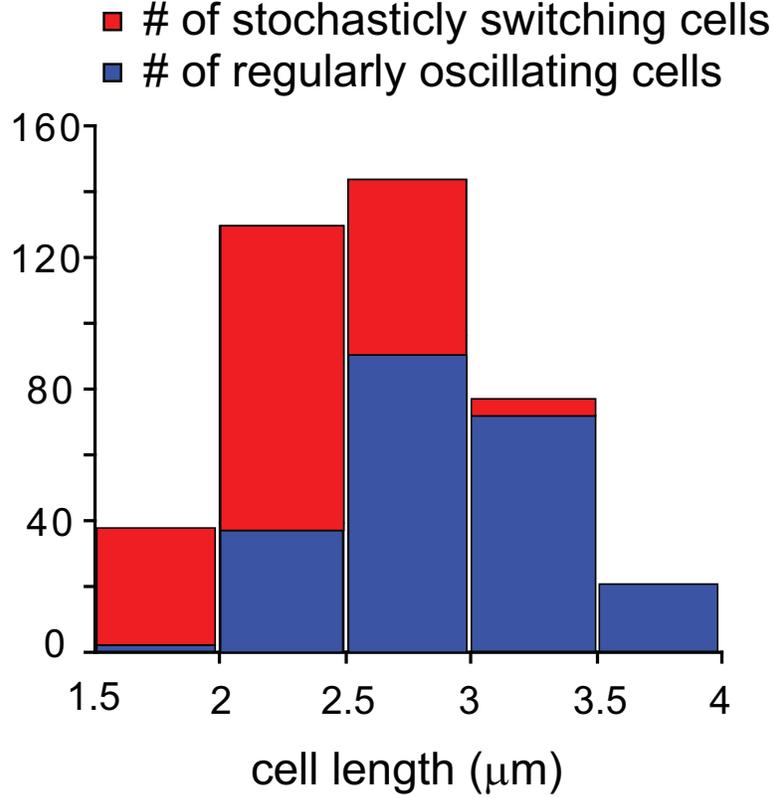}
\caption{Histogram of Min switching behavior of cells in a given cell length interval. The red column counts stochastically switching cells whereas the blue column represents the regularly switching cells. The decision whether a cell switches stochastically or not was taken as follows: residence times from a cell where sorted according to the cell length interval which the cell was in  at the beginning of the residence time. If there were more than 3 residence times associated to a given cell length interval, with  mean smaller than $80$s, than we calculated $r_{\rm ic}=\sigma_{\rm ic}/\bar\tau_{\rm ic}$, where  $\bar\tau_{\rm ic}$ is the mean and $\sigma_{\rm ic}$ the standard deviation. If this quantity was greater than one, the cell was decided to switch stochastically in this cell length interval.
If there were less than 3 residence times in the corresponding length interval or if the mean  $\bar\tau_{\rm ic}$ was greater than $80$s, then we looked at the maximal residence time occurring. If it was greater than $100$s, we labeled the cell as stochastically switching in the respective length interval.}
      \label{supp1}
   \end{center}
\end{figure}

We analyzed the stochasticity of MinD switching in
individual cells for time intervals in which the cell grew less than $\Delta l$. For the associated Min dynamics in such an interval, we calculated the ratio $r_{\rm ic}$ of the standard deviation $\sigma_{\rm ic}$ of the residence times and the corresponding mean $\bar\tau_{\rm ic}$,
$$
r_{\rm ic}=\sigma_{\rm ic}/\bar\tau_{\rm ic}.
$$
This quantity measures the stochasticity of the Min dynamics in an individual cell -- the higher it is, the more stochastic is the switching behavior. In Fig.~\ref{supp1}, we show a histogram of the number of cells which  switch  stochastically/regularly in a length interval of $\Delta l=0.5\mu$m according to the stochasticity  measure $r_{\rm ic}$.

\begin{figure}
   \begin{center}
      \includegraphics*[angle=0,width=5in]{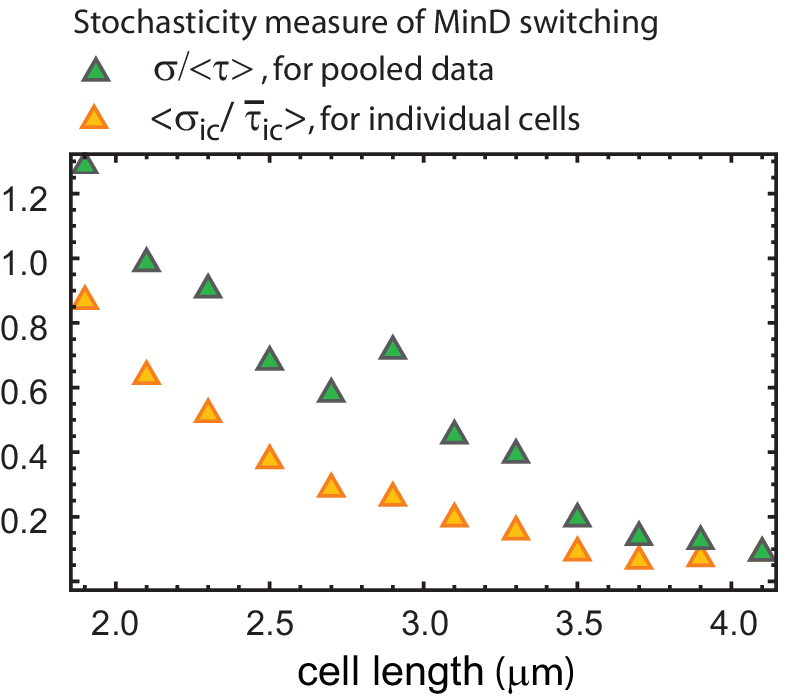}
\caption{The evolution of stochasticity of Min dynamics with increasing cell length.
Yellow triangles: population average of $r_{\rm ic}$,
where $r_{\rm ic}$ denotes the ratio of the standard deviation to the mean  of residence times for an individual cell in a time intervall in which the cell grew less than $0.2 \mu \rm m$, see text for details.
Green triangles: The same data of residence times were first pooled for cells in the same length interval. Then, ratio $r=\sigma/\langle \tau\rangle$ of the standard deviation to the mean of these residence times were calculated.}
      \label{supp2}
   \end{center}
\end{figure}

We averaged the quantity $r_{\rm ic}$ over ensembles of cells in a cell length interval $\Delta l=0.2\mu$m.
In Fig.~\ref{supp2}, we present the average ratio $\langle r_{\rm ic}\rangle$
as a function of the cell length, (yellow triangles), where $\langle\ldots\rangle$ denotes
the average over all measured cells in the respective length interval. To get a meaningful estimate of $r_{\rm ic}$, only cells
showing at least 5 switching events within a length interval of $0.2\mu$m were considered. Very long residence
times are thus neglected, which, together with the small number of events entering $r_{\rm ic}$, tends to
systematically underestimate the standard deviation of the "real" distribution. The value of $r$ initially drops
monotonically and stays constant for cell lengths larger than 3.5$\mu$m. The ratio of the mean residence time
and the corresponding standard deviation decreases thus in individual cells.

For comparison, we took the residence times of the same cells and pooled them directly according to the appropriate cell length interval. This means that the averaging over cell populations is done first, before any statistical analysis.
From the obtained sets of residence time distributions, we calculated again standard deviation $\sigma$ and mean $\langle \tau\rangle$ and their ratio $r=\sigma/\langle \tau\rangle$. Again, this quantity $r$ drops for increasing cell length, Fig.~\ref{supp2} (green triangles), but takes on average higher values than $\langle r_{\rm ic}\rangle$ (yellow triangles). This is because  the standard deviation results in this case also from cell-cell variability.

\subsection{Regularly Oscillating Cells, Protein Translocation Events Are Occasionally Missed}
\begin{figure}
   \begin{center}
      \includegraphics*[angle=0,width=5.25in]{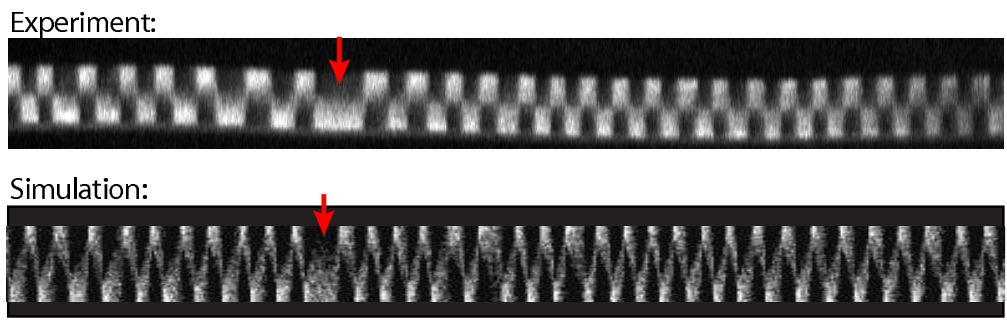}
\caption{Examples of missed switching events in otherwise regularly oscillating cells. The top panel shows a kymograph from a real cell. The bottom panel was obtained from a simulation.}
      \label{supp3}
   \end{center}
\end{figure}

Rarely, in regularly oscillating cells, a switching event
is missed and the fluorescence maximum stays longer in one cell
half. An example of such an event is shown in the top panel of
Fig.~\ref{supp3}. This phenomenon can also be observed in the simulated
Min oscillations as shown in the bottom panel of Fig.~\ref{supp3}. These
events give rise to a small tail in the residence time distribution of
oscillating cells.

\subsection{MinE Ring Stochastically Switches in Short Cells}
\begin{figure}
   \begin{center}
      \includegraphics*[angle=0,width=5.25in]{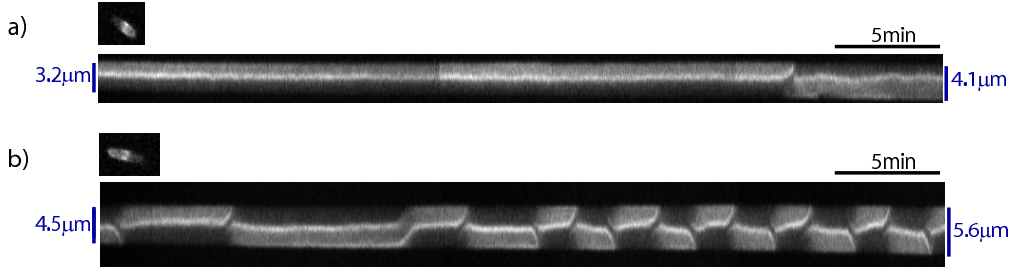}
\caption{Kymograph of stochastic switching of fluorescent labeled MinE in two cells of the {\it E.coli} strain WM1079. The kymographs cover a
time span of 40 minutes.}
      \label{supp4}
   \end{center}
\end{figure}

During our measurements
of MinD fluorescence in cells, we could not assess
the distribution of MinE simultaneously. Instead, we observed
the distribution of fluorescently labeled MinE in a different
Escherichia coli strain, WM1079 (S1). This strain produces on
average larger cells and does not show a clear transition from stochastic
Min switching to Min oscillations during cell growth. We
did however observe stochastic switching of MinE in individual
cells, Fig.~\ref{supp4}. This fact together with known results of simultaneously
recorded distributions of MinD and MinE in vivo (S2)
and in vitro (S3) suggests strongly that MinE is in the same
way distributed in stochastically switching cells as in normally
oscillating cells: A MinE ring at the rim of the MinD tube and
a shallow MinE layer on the remainder of the MinD tube.

\subsection{MinD is Partitioned Unevenly Between the Daughter Cells During Cell Division}
\begin{figure}
   \begin{center}
      \includegraphics*[angle=0,width=4.in]{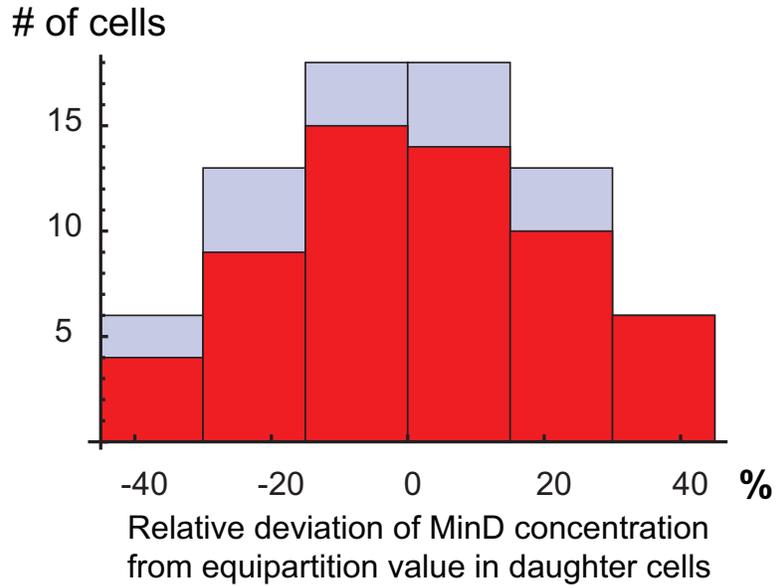}
\caption{Histogram of the deviation from the equipartition value $I_{eq}=(I_{int}^{D1}+I_{int}^{D2})/2$ of the integrated MinD fluorescence
$I_{int}^D$ in a daughter cell  after cell division. Given is the relative deviation $I_{int}^D/I_{eq}$ in percent.
The red part of the columns indicates the number of daughter cells which switched stochastically after division and the blue part of the columns those which oscillated regularly. Note that, since we assume that no GFP-MinD are lost during division, the overall histogram has to be symmetric around zero.}
      \label{supp5}
   \end{center}
\end{figure}

We measured the total fluorescence intensity from
daughter cells after cell division (the estimated error is about
10\%). We then calculated the associated deviation of the daughter
fluorescence from the equipartition value, i.e., the total fluorescence
intensity which would be expected if MinD was partitioned
equally to both daughter cells. Data from 37 cell divisions
give a distribution of relative deviations from the equipartition
value with a standard deviation of 21\%, Fig.~\ref{supp5}. The distribution
peaks in the center meaning that a zero deviation is most likely.
This is in contrast to the Min distribution suggested by the
theoretical results presented by Tostevin and Howard (S4).
There, the authors probed the partitioning of Min proteins using
a one-dimensional stochastic model including MinD polymerization
and the formation of MinD/MinE complexes on the membrane.
They simulated the division process assuming a gradually
decreased diffusion constant in the cell middle due to cell constriction.
As a result, they predicted a distribution of MinD
fractions in the daughter cells which peaked at 20\% deviation
from the equipartition value.

\subsection{Increasing MinE Attachment Rate can Generate a Transition from
Stochastic Switching to Regular Oscillations in Cells of Fixed Length}
\begin{figure}
   \begin{center}
      \includegraphics*[angle=0,width=4.in]{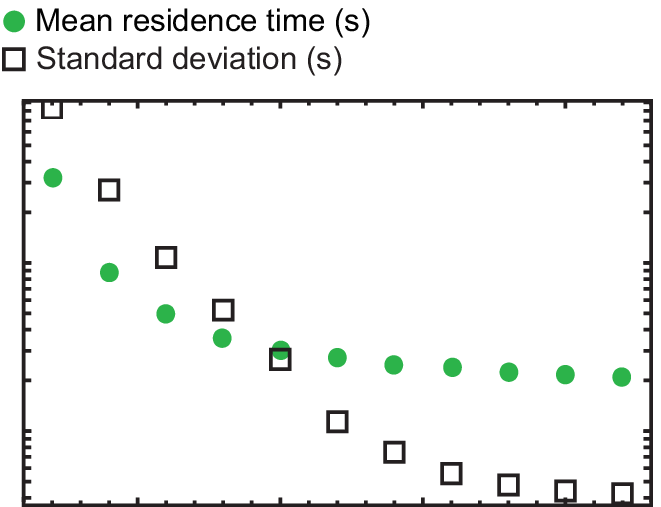}
\caption{Mean residence time (green dots) and standard deviation (black squares) in a simulated cell for increasing attachment rate $\omega_E$ of MinE but at constant cell length. The MinD/MinE ratio is kept constant as 8/3. Standard deviation and residence time both fall for increasing concentrations. Their ratio also decreases, implicating reduced stochasticity of Min dynamics.}
      \label{supp6}
   \end{center}
\end{figure}

We tested how the Min-protein dynamics changes within our stochastic one-dimensional model with an increasing MinE attachment rate $\omega_E$ in cells of a \textit{fixed} length of $2.5\mu$m.
We chose a mean MinD protein number of 1800 and a fixed MinD/MinE ratio of $8/3$. For each value of $\omega_E$ we performed 70 simulations with simulation time $160$min each. To mimic the variation of actual protein numbers in real cells, we drew the protein numbers of MinD and MinE from a Gaussian distribution with 10\% standard deviation.
The remaining parameters of the simulation were $\omega_D=0.04s^{-1}$,
$\omega_{de}=0.04s^{-1}$, $D_d=0.06\mu {\rm m}^2/s$, $r_d=1.2\mu{\rm
m}$, $r_{de}=0.1\mu{\rm m}$,
$g_d=35k_B T$, $g_{de}=-20k_B T$, $n_{\rm max}=43$ and bin length $l_b=33$nm. As a result, we find that the stochasticity of the switching is reduced for increasing $\omega_E$ coupled with a drop in mean residence time, Fig.~\ref{supp6}. For $\omega_E>0.3s^{-1}$ we observe regular oscillations.

\subsection{Stochastic Model Switching Can be Triggered by Other Parameters
Than $\omega_E$.}
As mentioned in the main text and in the previous paragraph, our model predicts a transition from stochastic Min switching to regular oscillations by an increase of the binding rate of MinE to the cytoplasmic membrane.
In addition, other model parameters  can trigger such a transition in the dynamics (S5). It can be induced by
i) an increase of the MinE/MinD ratio in the cell, by  ii) a decrease in the density of binding sites on the membrane $n_{max}$, by  iii)  a joint increase of MinE and MinD concentrations, and by iv) a change in MinD binding or MinD/MinE detachment from the membrane (data not shown). A pure increase of the cell length keeping the remaining system parameters constant (including protein concentrations, not protein numbers)  results in a trend towards more stochastic switching for longer cells. This is the contrary effect to what we observe experimentally. This effect can be (over)compensated though by an increase of e.g. $\omega_E$, as has been shown in Fig.~\ref{fig5} (main text). Only for very short cells smaller than $1.9\mu$m, the model predictions deviate from the experimental data for the parameter choice presented in Fig.~5. In this length regime, it predicts a decrease of the mean residence time towards smaller cell lengths.

\subsection{Analysis of Alternative Models with Regard to Stochastic Switching}
We also analyzed stochastic versions of models presented in
(S6) and in (S3) with regard to stationary states and stochastic
switching in small E. coli cells.
The stochastic simulations that we performed were one-dimensional
and relied on a particle based description together
with spatial binning ($\Delta x =$0.13$\mu$m). The reaction dynamics and
particle exchange between the spatial bins due to diffusion was
implemented with a Gillespie algorithm (S8). The rates for the
attachment, detachment and diffusion processes were chosen
as indicated in the refs. S3, S6, S7.

In these models, we did not find switching behavior similar to
what we observed in cells. We systematically scanned the neighborhood
of the parameter space of the models given in (S6, S7)
and (S3), respectively, including independent increase of the
parameters up to a factor of 2 and decline up to a factor of
0.5. Particle numbers were varied in the simulations up to
20\% from the originally given values.We did not chose a broader
range, because the particle densities are chosen according to experimental
findings (S2).We always tested cell lengths between 2
and 6 microns. The solutions we found for the models presented
in (S3, S6, S7) were either oscillations or noisy homogeneous
states.

\clearpage

\end{document}